\documentclass[pra,twocolumn,showpacs,floatfix,amsfonts]{revtex4}
\usepackage{bm}
\usepackage{graphicx}
\usepackage{latexsym}                
\usepackage{amsmath}
\usepackage{amsfonts}
\usepackage{amssymb}
\usepackage{dcolumn}
\usepackage{color}

\usepackage{amsthm,graphics,epsfig}

\theoremstyle{plain}
\newtheorem{theorem}{Theorem}

\theoremstyle{definition}
\newtheorem{definition}[theorem]{Definition}

\newcommand{\bra}[1]{\langle #1 | \,}
\newcommand{\ket}[1]{\, | #1 \rangle}
\newcommand{\braket}[2]{\langle #1 | #2 \rangle}
\newcommand{\ketbra}[2]{| #1 \rangle\langle #2 |}

\newcommand{\be}{\begin{equation}}
\newcommand{\ee}{\end{equation}}
\newcommand{\bea}{\begin{eqnarray}}
\newcommand{\eea}{\end{eqnarray}}

\newcommand{\N}{{N}}
\begin{document}
\title{Deterministic quantum-public-key encryption: forward search attack and randomization}
\author{Georgios M. Nikolopoulos}
\affiliation{Institute of Electronic Structure {\it \&} Laser, FORTH, P.O. Box 1527, Heraklion 71110, Crete, Greece}
\author{Lawrence M. Ioannou}
\affiliation{Center for Quantum Computation, University of Cambridge, Wilberforce Road,
Cambridge, CB3 0WA, UK}

\date{\today}

\begin{abstract}
In the classical setting, public-key encryption requires randomness
in order to be secure against a forward search attack, whereby an
adversary compares the encryption of a guess of the secret message
with that of the actual secret message. We show that this is also
true in the information-theoretic setting --- where the public keys
are quantum systems --- by defining and giving an example of a
forward search attack for any deterministic quantum-public-key
bit-encryption scheme. However, unlike in the classical setting, we
show that any such deterministic scheme can be used as a black box
to build a randomized bit-encryption scheme that is no longer
susceptible to this attack.
\end{abstract}


\maketitle

\section{Introduction}
Quantum-public-key cryptography, where the public keys are
quantum-mechanical systems, was introduced by Gottesman and Chuang
in Ref. \cite{qphGC01}, which contains an information-theoretically
secure quantum digital signature scheme for signing classical
messages.  Other explorations within this information-theoretic
framework include a no-go theorem for signing arbitrary quantum
states \cite{BCGST02}, ``lock and key'' systems and distribution of
quantum public keys \cite{ACJ06}, identification schemes
\cite{qphIM08}, and --- our focus in this paper --- encryption schemes
\cite{Got05,KKNY05,HKK08,Kak06,Nik08}.

Roughly put, the purpose of an encryption scheme is to facilitate
the communication of some secret information over an insecure
channel, from a sender to a receiver, such that an adversary, who
has access to this channel, cannot obtain anything close to a
meaningful representation of the secret information. This secret
information is called the \emph{plaintext}, while the actual signal
sent over the channel, which somehow \emph{encodes} the plaintext,
is called the \emph{ciphertext}. In the classical setting,
public-key encryption requires randomness in order to be secure
against a forward search attack, whereby an adversary compares the
ciphertext encoding a guess of the plaintext or ---
\emph{test-plaintext} --- with the ciphertext she is trying to
decrypt (see Ref. \cite{Gol04} for more details in the classical
setting). We show that this is also true in our
information-theoretic setting (defined in Section \ref{secII}), by
defining and giving an example of a forward search attack for any
deterministic bit-encryption scheme that uses quantum public keys.
However, unlike in the classical setting, we show that any such
deterministic scheme can be used as a black box to build a
randomized bit-encryption scheme that is no longer susceptible to
this attack.

\section{Quantum-public-key encryption} 
\label{secII}

The potential for information-theoretic security in the
quantum-public-key setting arises from the existence of a quantum
function, mapping classical private keys (binary strings) to
corresponding quantum public keys (quantum-mechanical systems), that
is impossible to invert.  More precisely, we have the following
general setup.  All users of the cryptosystem agree on a classical
description of a set
\begin{eqnarray}
A(n) \equiv \{\ket{\Psi_x}: x \in \{0,1\}^n\}
\end{eqnarray}
of $\log_2(d)$-qubit pure states (in general, $d=d(n)$) such that,
for any distinct $x$ and $x'$ in $\{0,1\}^n$,
\begin{eqnarray}
|\braket{\Psi_{x'}}{\Psi_x}| < \delta
\end{eqnarray}
for some positive constant $\delta < 1$.  Any user can now choose a
uniformly random \emph{private key} $k \in \{0,1\}^{n}$ and then
generate and distribute (at most) $T$ quantum-mechanical systems in
or --- \emph{copies of} --- the state $\ket{\Psi_k}$; each copy of
$\ket{\Psi_k}$ constitutes one \emph{(quantum) public key}.  We
assume that each public key reaches its intended recipient in an
authenticated fashion.  The bijective map
\begin{eqnarray}
x\mapsto (\textrm{$T$ copies of $\ket{\Psi_{ x}}$})
\end{eqnarray}
is a one-way (quantum) function in the sense that, for a given
${x}\in \{0,1\}^n$, the deterministic preparation of a system in the state
$\ket{\Psi_x}$ is possible via the classical description of $A(n)$,
while the inversion of the map (with nonnegligible probability) is
guaranteed impossible by the Holevo bound \cite{Hol73} when \be
\label{bound_Holevo}n \gg T\log_2(d). \ee This inequality thus sets
an upper bound on the number $T$ of public keys that can be publicly
distributed, in order to ensure the secrecy of the private key,
which is the minimal requirement for security of any cryptographic
scheme in this framework.  Note that the notion of computational
efficiency may be ignored in an information-theoretic setting;
however, there do exist constructions of $A(n)$ such that $n$ is
large enough that the set is cryptographically useful and such that,
for all $x \in \{0,1\}^n$, a copy of $\ket{\Psi_x}$ can be computed
in (quantum-) polynomial time from input $x$ \cite{BCWW01, qphGC01}.

Within the above framework, a \emph{deterministic}
quantum-public-key bit-encryption scheme may be defined by further
specifying (and publishing, along with the description of $A(n)$)
two unitary encryption operators, $\hat{\mathcal{U}}_0$ and
$\hat{\mathcal{U}}_1$, and a decryption procedure whose exact form
does not concern us.  If Bob wants to communicate the plaintext $b
\in \{0,1\}$ to Alice, he obtains an authenticated copy of Alice's
public key, which is, by definition, in the state $\ket{\Psi_k}$,
creates the (quantum) ciphertext in the state $\ket{\Phi_{k,b}}
\equiv \hat{\cal U}_b \ket{\Psi_k}$, and sends it to Alice, who then
decrypts and recovers the plaintext. Note that $\hat{\mathcal{U}}_0$
and $\hat{\mathcal{U}}_1$ do not depend on the private key $k$, but
Alice's decryption procedure does.

Of course, in general, in our quantum setting, the plaintext can
also be quantum, i.e., it can be a quantum-mechanical system in a
particular state.  Thus, we are focussing on the case where (a
classical description of) the set of all possible (quantum)
plaintexts consists of just two orthogonal states, $\ket{0}$ and
$\ket{1}$. This is in fact the most general case from a security
point of view: it may be seen as corresponding to the case where the
adversary has narrowed down the plaintext to one of two
maximally-distinguishable possibilities (of course, the states of
the corresponding ciphertexts need not be orthogonal, depending on
the encryption scheme; but, in any reasonable scheme, orthogonal
plaintext-states would give rise to maximally-distinguishable
ciphertexts, for a given key-value).  However, we do not formally
define what it means for an encryption scheme to be secure, because
we do not prove security of any scheme; we only ever refer to
security against a particular attack, i.e., our forward search
attack.

In the following, we may abuse terminology by referring to quantum
public keys or ciphertexts by their classical descriptions, i.e., by
their states.

\section{Forward search attack based on a symmetry test}
\label{secIII}

Before defining ``(quantum) forward search attack'', we should
remind ourselves of what is the most general attack for uncovering
the plaintext encoded by a particular ciphertext (as opposed to an
attack that tries to compute the private key). If an adversary, Eve,
wants to decide what the plaintext $b$ is, given the ciphertext
$\ket{\Phi_{k,b}}$ and all $(T-1)$ possible copies of the public key
$\ket{\Psi_{k}}$, then she is ultimately faced with the problem of
deciding which of the following two states she has:
\begin{eqnarray}
\rho_0 &\equiv& \frac{1}{2^n}\sum_{x}\ketbra{\Psi_{x}}{\Psi_{x}}^{\otimes (T-1)}\ketbra{\Phi_{x,0}}{\Phi_{x,0}}\\
\rho_1 &\equiv&
\frac{1}{2^n}\sum_{x}\ketbra{\Psi_{x}}{\Psi_{x}}^{\otimes
(T-1)}\ketbra{\Phi_{x,1}}{\Phi_{x,1}}.
\end{eqnarray}
The optimal procedure (``POVM'') for solving this ``binary quantum
decision problem'' is given in Refs. \cite{Hel76, Fuc95} and depends
on $\rho_0$, $\rho_1$, and the prior probability distribution
$(p,1-p)$ of the plaintext $b$ (i.e. $P[b=0]=p$).  We assume that
Eve can implement this optimal procedure, since we do not place any
computational resource-bounds on her.  The probability of success of
this optimal procedure, which is affinely related to the trace
distance between $p\rho_0$ and $(1-p)\rho_1$, is in general
difficult to calculate.

In this paper, we concentrate on a restricted class of attacks that
attempt to uncover the plaintext encoded by a particular ciphertext.
\begin{definition}[Forward search attack]
A \emph{forward search attack} on a deterministic quantum-public-key
bit-encryption scheme is any (quantum) algorithm --- independent of
the encryption and decryption operations and the structure of the set of public
keys --- that outputs the plaintext with some probability of error,
given one copy of the actual ciphertext and all available copies of
the ciphertext encoding a test-plaintext.
\end{definition}
\noindent As an aside, we note that this definition subsumes the
definition of ``forward search attack'' for computationally-secure,
classical public-key bit-encryption schemes that are implemented
quantum-mechanically \footnote{For this to be true, the only
assumption needed is that the set of ciphertexts of any such scheme
is a subset of the computational basis, so that the outcome of a
joint measurement with respect to the computational basis of the
actual ciphertext and one copy of the test-ciphertext determines
with certainty whether the two ciphertexts are identical.}.  In the
following, we give a simple forward search attack that we suspect is
near to the optimal forward search attack and whose probability of
success is easily computed. To simplify our presentation, we assume
that each plaintext is equally likely and thus always use the
test-plaintext 0 without loss of generality.

\begin{figure} [htbp]
\vspace*{13pt}
\centerline{\epsfig{file=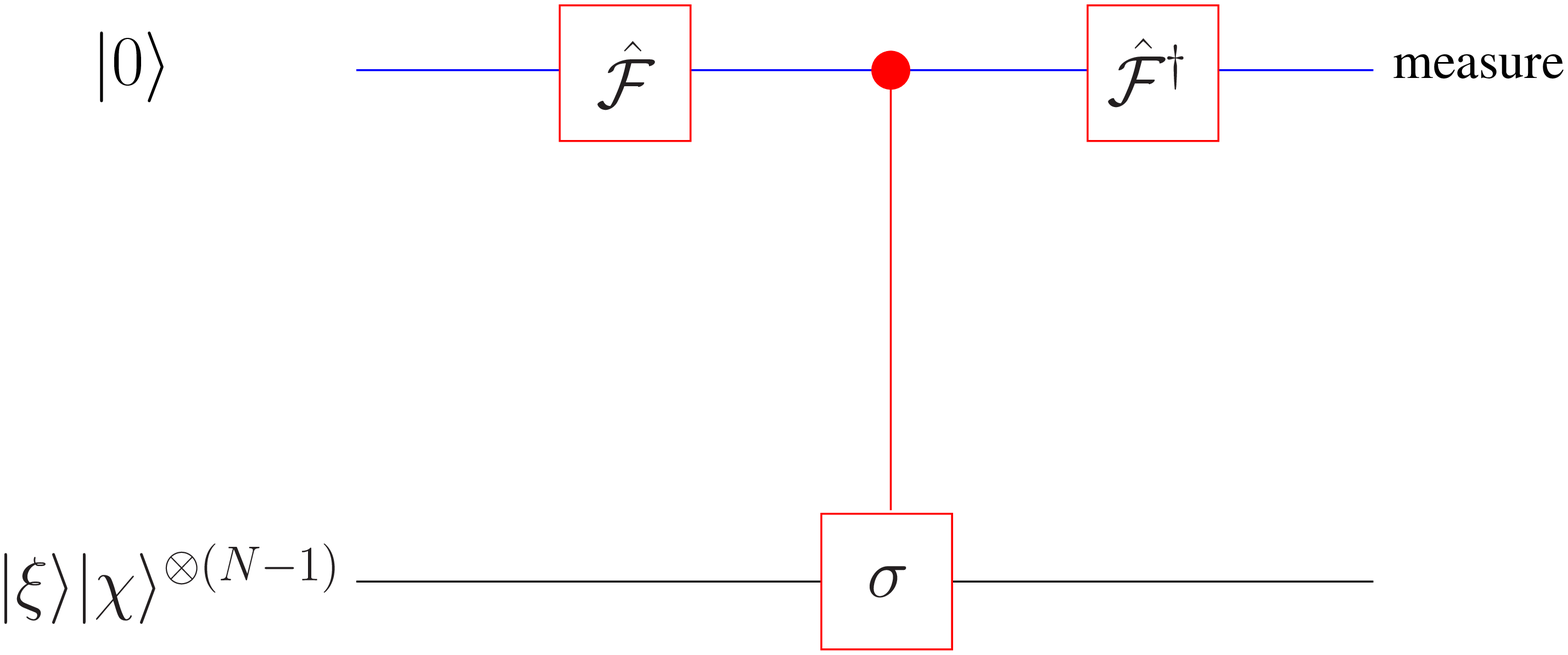, width=8.2cm}} 
\vspace*{13pt} \caption{(Color online) Symmetry-test for the
$(1,\N-1)$-copy state distinguishing problem.  The top (blue) wire
is a $\N!$-dimensional quantum system, whose state-space is spanned
by the computational basis states, each of which is labeled by a
permutation $\sigma \in S_\N$ (e.g. $\ket{0}$ corresponds to the
identity permutation). The bottom wire represents $\N$ registers,
each of dimension $d$.}\label{permtest}
\end{figure}

Following Ref. \cite{BCWW01}, we first define a problem that
captures the essence of Eve's task of determining the plaintext
via forward search attack (i.e. ignoring all structure of the
particular cryptosystem), and then we give a solution for it, based
on a test for symmetry.
\begin{definition}[$(1,\N-1)$-copy state distinguishing problem]
Given one copy of $\ket{\xi}\in \mathbb{C}^d$ and $(\N-1)$ copies of
$\ket{\chi}\in \mathbb{C}^d$ such that either $\ket{\xi}=\ket{\chi}$
or $|\braket{\xi}{\chi}| = \lambda <1$, decide which case holds.
\end{definition}
\noindent To solve this problem with some probability of error, we
can use the \emph{symmetry-test} procedure depicted in Fig.
\ref{permtest}, which we now explain.
Let $S_\N$ be the set of all $\N!$ permutations on $\N$ objects and
let $\sigma \in S_\N$.  The operator $\hat{\mathcal{F}}$ is the
$(\N!)$-dimensional quantum Fourier transform \cite{BCWW01}, so
that, in particular,
\begin{eqnarray}
\hat{\mathcal{F}}\ket{0} = \frac{1}{\sqrt{\N!}}\sum_{\sigma \in
S_\N} \ket{\sigma},
\end{eqnarray}
and the controlled-$\sigma$ operator permutes the $\N$ target
registers according to the permutation $\sigma$ encoded by the
computational-basis-state of the control register. The probability
of the final measurement in the computational basis of the top
register resulting in outcome ``0'' is 1 when $\ket{\xi} =
\ket{\chi}$.  But when $\braket{\xi}{\chi} = \lambda <1$, this
probability is
\begin{subequations}
\label{pfail}
\begin{eqnarray}
&&\left\|(\ketbra{0}{0}\otimes I) \frac{1}{\sqrt{\N!}} \sum_{\sigma \in S_\N} \hat{\mathcal{F}}^{\dagger} \ket{\sigma} \sigma(\ket{\xi}\ket{\chi}^{\otimes (\N-1)}) \right\|^2\hspace{9mm}
\label{pfal_a}\\
&=& \left\|\frac{1}{\sqrt{\N!}} \sum_{\sigma \in S_\N} \bra{0}\hat{\mathcal{F}}^{\dagger} \ket{\sigma} \sigma(\ket{\xi}\ket{\chi}^{\otimes (\N-1)}) \right\|^2\nonumber\\
&=& \left\|\frac{1}{\N!} \sum_{\sigma \in S_\N} \sigma(\ket{\xi}\ket{\chi}^{\otimes (\N-1)}) \right\|^2 \nonumber\\
&=&\frac{1}{\N!^2} \sum_{\sigma, \tau \in S_\N}\tau(\bra{\xi}\bra{\chi}^{\N-1})\sigma(\ket{\xi}\ket{\chi}^{\N-1})\nonumber\\
&=&\frac{1}{\N!^2} \sum_{\sigma, \tau \in S_\N}\bra{\xi}\bra{\chi}^{\N-1}\tau^{-1}\sigma(\ket{\xi}\ket{\chi}^{\N-1})\nonumber\\
&=&\frac{1}{\N!} \sum_{\sigma \in S_\N}\bra{\xi}\bra{\chi}^{\N-1}
\sigma(\ket{\xi}\ket{\chi}^{\N-1})\nonumber\\
&=& \frac{(\N-1)!(1!)}{\N!} \left({\N-1 \choose 0}(1) + {\N-1 \choose 1}(|\braket{\xi}{\chi}|^2)\right)\nonumber\\
&=& \frac{1}{\N} (1 + (\N-1)\lambda^2)\label{pfal_b},
\end{eqnarray}
\end{subequations}
which we denote $q_{\N,\lambda}$.  Thus, we only care whether the
measurement outcome is ``0'' or not: in case it is ``0'', we guess
that $\ket{\xi}=\ket{\chi}$ (but we might be wrong); otherwise, we
know that $\braket{\xi}{\chi} = \lambda <1$.  With this strategy, we
can only make an error when $\braket{\xi}{\chi} = \lambda <1$, in
which case the error probability is $q_{\N,\lambda}$.

Thus, to perform a \emph{forward search attack by symmetry-test},
Eve applies the above procedure (and decision strategy), with
\begin{eqnarray}
\ket{\xi} &\equiv& \ket{\Phi_{k,b}},\\
\ket{\chi} &\equiv& \ket{\Phi_{k,0}},
\end{eqnarray}
\noindent and the maximum possible $\N$.  For a (non-classical)
quantum-public-key bit-encryption scheme, Eve can use $\N = T$, thus
obtaining one-sided error $q_{T,\lambda}$ \footnote{For a classical,
computationally-secure scheme implemented quantum-mechanically, Eve
can use arbitrarily large $\N$ so that her error is arbitrarily
close to zero, as we would expect.}.  Although we only suspect that this
forward search attack is nearly the optimal one, we note that the 
same symmetry-test procedure is nearly optimal for the ``$(\N',\N')$-copy 
state distinguishing problem'', where one is
given $\N'$ copies each of $\ket{\xi}$ and $\ket{\chi}$ (and the
procedure permutes $2\N'$, instead of $\N$, target registers)
\cite{BCWW01}. In the remainder of this work, we show that our
assumption that Eve's probability of correctly guessing the
plaintext (by forward search attack) is bounded away from 1 leads to
a simple randomized encryption scheme that uses the original
deterministic scheme as a black box and is resistant to our forward
search attack.

\section{Randomization against forward search attack}
\label{secIV}

Any deterministic public-key bit-encryption scheme, quantum or
classical, is susceptible to a forward search attack.  However, if
the scheme can be \emph{nontrivially extended} to encrypting
multiple-bit plaintexts --- by which we mean that the multiple-bit
scheme is not merely the concatenation of instances of the original
single-bit scheme --- one possible way to guard against a forward
search attack is to use the following \emph{parity encoding}. If the
desired plaintext is $b \in \{0,1\}$, Bob should first choose a
uniformly random, binary-string \emph{codeword} $w$, whose length is
$s > 1$ and whose (Hamming) weight (sum of the bits) has parity $b$,
and then encrypt $b$ by using the $s$-bit version of the
deterministic scheme to encrypt $w$, i.e., the new ciphertext
encoding $b$ is actually the ciphertext encoding $w$. Assuming Alice
knows that the intended plaintext $b$ is actually the parity of the
weight of $w$, then this forms a randomized bit-encryption scheme
that, for sufficiently large $s$, may not be susceptible to the
forward search attack (of course, we do not claim that the use of
the parity encoding results in a secure bit-encryption scheme, in
general). The parameter $s$ thus functions as a ``security
parameter''.

Now consider the case where the original deterministic
bit-encryption scheme has no nontrivial extension to multiple-bit
plaintexts. Can it be used several times (under different
key-values) as a black box, in order to create a randomized scheme
that is potentially secure against a \emph{compound forward search
attack}, whereby Eve does a forward search attack on every instance
of the original scheme?  In the classical setting, the answer is
clearly ``no'': Eve would learn the correct plaintext in every
instance of the original scheme, so Alice would have no advantage
over her. In our quantum setting, however, the answer to this
question is ``yes'', as shown by the following randomized
bit-encryption scheme, which just combines the above parity encoding
with the trivial multiple-bit extension of the original scheme.
Assume that Alice's public key is now $\otimes_{i=1}^{s}
\ket{\Psi_{k_i}}$, where each $k_i$ is uniformly randomly chosen
from $\{0,1\}^n$.  To encrypt plaintext $b\in \{0,1\}$, Bob again
first chooses a uniformly random codeword $w$, whose length is $s >
1$ and whose weight has parity $b$. The ciphertext that encodes $b$
is now simply $\otimes_{i=1}^s \ket{\Phi_{k_i , w_i}}$, where $w =
w_1w_2\cdots w_s$.  Alice decrypts to get $w$, and thus the intended
plaintext $b$ as the parity of the weight of $w$.

Consider Eve's compound forward search attack by symmetry-test on
this new scheme, whereby Eve does $s$ separate forward search
attacks by symmetry-test as described in the previous section, one
for each value of $i$.  We now assume that distinct ciphertexts
(under the same key-value) in the original bit-encryption scheme are
orthogonal, i.e.,
$\lambda\equiv\braket{\Phi_{k_i,0}}{\Phi_{k_i,1}}=0$ for all $i$
(this restricts to schemes where decryption is perfect).  Assuming
Eve uses $\ket{\chi} = \ket{\Phi_{k_i,0}}$ for all $i$, she can only
fail in guessing $w_i$ correctly when $w_i = 1$.  Each codeword $w$
has a weight $\alpha$ of well defined parity. Thus, a codeword will
be decrypted correctly if, for some even $\gamma \in \{0,1,\ldots,
\alpha\}$, $\gamma$ out of $\alpha$ symmetry-tests give measurement
outcome ``0'' and $(\alpha-\gamma)$ symmetry-tests give a different
outcome. On average, the probabilities for Eve to decrypt
successfully each of the bit values are
\begin{widetext}
\bea P^{(s)}(\textrm{success}|b=0)&=&
\frac{1}{2^{s-1}}\sum^{s}_{\substack{\alpha=0 \\
    \textrm{even}}}~\sum^{\alpha}_{\substack{\gamma=0 \\
    \textrm{even}}}\binom{s}{\alpha}
\binom{\alpha}{\gamma}q^\gamma(1-q)^{\alpha-\gamma},\\
P^{(s)}({\textrm{success}}|b=1)&=&\frac{1}{2^{s-1}}\sum^{s}_{\substack{\alpha=1 \\
    \textrm{odd}}}~\sum^{\alpha}_{\substack{\gamma=0 \\
    \textrm{even}}}\binom{s}{\alpha}
\binom{\alpha}{\gamma}q^\gamma(1-q)^{\alpha-\gamma},
\eea
\end{widetext}
where $q \equiv q_{T,0}$. Since we assume both plaintexts are
equally probable, we have
\begin{eqnarray}
&&\hspace*{-0.5cm}P^{(s)}({\textrm{success}})\nonumber\\
&=&\frac{1}{2}\left [ P^{(s)}(\textrm{success}|b=0)
+P^{(s)}({\textrm{success}}|b=1)\right ]\nonumber\\
&=&\frac{1}{2}+\frac{(1-q )^s}{2}\label{lineB}\\
&=&\frac{1}{2}+\frac{(T-1)^s}{2T^s},
\end{eqnarray}
\noindent where the second-last line follows by mathematical
induction on $s$ \footnote{The proof consists of two steps. First,
it can be shown that Eq. (\ref{lineB}) holds for $s=1$, i.e.,
$P^{(1)}({\textrm{success}})=[1+(1-q)]/2$. Second, assuming that Eq.
(\ref{lineB}) holds for $s$, one can show that it also holds for
$s+1$. In this last step, one needs basic identities of binomial
coefficients, including $\sum^{n}_{j=0}\binom{s}{j}=2^n$ and
Pascal's rule $\binom{n}{j}+\binom{n}{j+1}=\binom{n+1}{j+1}$.}.~

Assume now that Alice and Bob have agreed in advance on a security
threshold $\epsilon\ll 1$, such that Eve's probability of success is
restricted to slightly above random guessing, i.e.,
$P^{(s)}({\textrm{success}})\leq 1/2+\epsilon$. This immediately
implies that the plaintext $b$ has to be encoded on
\begin{eqnarray}
s\geq\left |\frac{1+\log_2(\epsilon)}{\log_2\left (
\frac{T-1}{T}\right )}\right |
\end{eqnarray}
\noindent qubits.  Working on the right-hand side of this
inequality, we may derive a less tight, but simpler lower bound,
namely
\begin{eqnarray}
s\geq T|1+\log_2(\epsilon)|.
\end{eqnarray}
\noindent Assuming our forward search attack is the optimal one,
this condition is sufficient to thwart Eve's compound forward search
attack on the randomized bit-encryption scheme.

\section{Summary}

We have introduced the forward search attack in the framework of 
quantum-public-key encryption, which aims at recovering the plaintext 
from the ciphertext without reference to the structure of the particular 
encryption scheme. As in the classical public-key setting, any deterministic 
encryption scheme that uses quantum public keys is susceptible to such an attack, 
unless some sort of randomization is used.

Several quantum-public-key encryption schemes have been proposed,
the three most notable ones appearing in Refs.
\cite{Got05,KKNY05,Nik08}. 
The schemes in Refs. \cite{Got05,KKNY05} are randomized, with 
nontrivial extensions to multiple-bit plaintexts, and thus 
they are not vulnerable to a forward search attack   
\footnote{The mere fact that a (qu)bit-encryption scheme is
randomized is not necessarily enough for our forward search attack
to be ineffective: if the amount of randomness is dependent on (i.e.
limited by) the size of the plaintext, then the $s$-(qu)bit
extension of the scheme may have to be used in order to get a secure
bit-encryption scheme (even though the single-qubit-encryption
scheme may be secure for uniformly random qubit-plaintext with
respect to the Haar measure), by encoding the intended plaintext $b
\in \{0,1\}$ as the ciphertext that encodes the multi-qubit
plaintext $\ket{b}^{\otimes s}$, for some $s>1$.}. The scheme in
Ref. \cite{Nik08} is randomized in the way we have presented in Sec.
\ref{secIV}; our work places that scheme in the wider cryptographic
context.  In terms of computational efficiency, we note that the
schemes in Refs. \cite{Got05,KKNY05} require scalable quantum
computing in order to be secure against our forward search attack,
whereas the scheme in Ref. \cite{Nik08} requires only single-qubit
rotations about a fixed axis.

\section{Acknowledgements}
We would like to thank Daniel Gottesman for helpful discussions.
L. M. Ioannou acknowledges support from the EPSRC and SCALA.
G. M. Nikolopoulos acknowledges partial support from the EC RTN EMALI (contract No. MRTN-CT-2006-035369).


\begin{thebibliography}{99}
\bibitem{qphGC01} D.~Gottesman and I.~L.~Chuang, e-print arXiv:quant-ph/0105032.

\bibitem{BCGST02} H. Barnum, C. Cr{\'{e}}peau, D. Gottesman, A. Smith, and A. Tapp, in 
  {\em Proceedings of the 43rd Annual IEEE Symposium on the Foundations of Computer Science --- FOCS '02}, (IEEE Computer Society Press, Washington, DC, 2002) pp. 449-458.

\bibitem{ACJ06} E.~Andersson, M.~Curty, and I.~Jex, 
Phys. Rev. A {\bf 74}, 022304 (2006).

\bibitem{qphIM08} L.~M.~Ioannou and M.~Mosca, e-print arXiv:0810.2780.

\bibitem{Got05} D.~Gottesman, {\em Quantum public key cryptography with information-theoretic security}, Workshop on classical and quantum information security, Caltech, 15 - 18 December (2005), 
http://www.cpi.caltech.edu/quantum-security/program.html. 
See also http://www.perimeterinstitute.ca/personal/dgottesman/Public-key.ppt.

\bibitem{KKNY05} A.~Kawachi, T.~Koshiba, H.~Nishimura, and T.~Yamakami, in 
{\em Advances in Cryptology EUROCRYPT 2005}, 
Lect.  Notes Comput. Sci. Vol. 3494 (Springer, 2005), pp. 268-284. See also arXiv:quant-ph/0403069.

\bibitem{HKK08}
M.~Hayashi, A.~Kawachi, and H.~Kobayashi, Quantum Inf. Comput. {\bf 8}, 0345 (2008).

\bibitem{Kak06}
S.~Kak, Found. Phys. Lett. {\bf 19}, 293 (2006).

\bibitem{Nik08}
G.~M.~Nikolopoulos, Phys. Rev. A {\bf 77}, 032348 (2008); {\bf 78}, 019903 (2008).

\bibitem{Gol04}
O.~Goldreich, {\em Foundations of Cryptography} (Cambridge University Press, Cambridge, 2004), 
Vol. 2.  

\bibitem{Hol73}
A.~S.~Holevo, in {\em Proceedings of the Second Japan-USSR Symposium on Probability Theory}, 
edited by G. Maruyama and J. V. Prokhorov, Lect. Notes Math. Vol. 330 
(Springer-Verlag, Berlin, 1973), pp. 104-119.


\bibitem{BCWW01} H.~Buhrman, R.~Cleve, J.~Watrous, and R.~de Wolf, 
Phys. Rev. Lett. {\bf 87}, 167902 (2001).

\bibitem{Fuc95} C.~A.~Fuchs, PhD Thesis, University of New Mexico, 1995. 
See also e-print arXiv:quant-ph/9601020.

\bibitem{Hel76} C.~W.~Helstrom, {\em Quantum Detection and Estimation Theory}, Mathematics in 
Science and Engineering (Academic Press, 
New York, 1976), Vol. 123.

\end{thebibliography}
\end{document}